\documentclass{article}
\usepackage[utf8]{inputenc}
\usepackage{booktabs}
\usepackage{graphicx}
\usepackage{natbib}
\usepackage{authblk}
\usepackage{tabularx}
\usepackage{adjustbox}

\title{A Novel Approach to Participant-Level Influence Calculation in Viral Cascades}
\author[1,2]{Nick Hagar}
\author[1]{Laila Wahedi}
\author[1]{Eric Dunford}
\affil[1]{Meta Platforms, Inc., Menlo Park, CA}
\affil[2]{Northwestern University, Evanston, IL}
\date{February 2023}

\begin{document}

\maketitle

\abstract{Efforts to model viral cascades provide a vital view into how they form and spread. A range of methods, such as Multivariate Hawkes Processes or network inference algorithms, attempt to decompose cascades into constituent components via inference---by constructing an underlying network of influence, or by generating direct pairwise influence measures between cascade participants. While these approaches provide detailed models of the generative mechanics underlying event sequences, their sophistication often comes at a steep computational cost that prevents them from being applied to large-scale datasets. This is particularly the case for Multivariate Hawkes Processes. In this work, we propose a novel, scalable method for generating individual-level influence measures across a set of cascades. Across real-world datasets, we demonstrate the alignment of this approach's calculations with the influence inferred by established methods, as well as the computational scalability of this method.}

\section{Introduction}
Event sequences provide a key structure for understanding time-dependent, ordered activity. By encoding information about the magnitude and spacing of a particular event, they allow detailed modeling of the underlying dynamics of activity and participation. Event sequences underpin a wide range of efforts to understand structured collective behavior, across domains like news consumption \citep{lerman2010using}, financial markets \citep{embrechts2011multivariate}, social influence \citep{luo2015inferring}, and others. 

As a conceptual model, sequences have proved a useful tool for examining the spread of digital content. The cascade model, in which successive spreading activity creates a large burst of engagement and exposure, has in particular been fruitful in this area of research \citep{cheng2014}. Cascades neatly map onto the observed dynamics of real-world sharing behavior, in which travel across social networks enables rapid exponential growth in popularity \citep{gleeson2016effects}. They also provide a flexible view of content virality, enabling analysis at a range of levels, from individual participation to dynamics across a set of content \citep{zhou2021survey}.

Examinations of cascades often focus on structure and its link to virality, or on popularity prediction at the level of specific items. For example, \citet{cheng2014,cheng2016cascades} both examine a range of cascade features to predict their eventual outcomes. \citet{lerman2010using} attempt to predict news popularity given information about user behavior and social network design. Similarly, \citet{szabo2010predicting} demonstrate the value that the momentum and trajectory of a piece of content provide in predicting its eventual popularity. These efforts are all concerned with predicting some popularity-related outcome from a given cascade, often with a wide range of engineered features \citep{zhou2021survey}.

In an alternative approach, researchers attempt to model cascades via the time and activity information encoded in event sequences. Rather than framing an explicit prediction problem, these approaches attempt to capture some high-level dynamics of the underlying historical data, either to project it forward or to explain some facet of the sequence's behavior. In this vein, \citet{vermeer2020online} use Markov chains to explore sequences of news consumption, in order to map out user journeys through a news site. More prevalent though are attempts to model the dynamics of cascades with point processes. This is the approach underlying SEISMIC, an attempt to model tweet popularity \citep{zhao2015seismic}, as well as many efforts involving Multivariate Hawkes Processes (MHPs) \citep{nickel2021modeling}.

This approach provides a valuable window into \textit{influence}. While often not directly measurable, influence allows us to express the extent to which an individual is consequential to how content spreads. Conceptually, influential individuals induce spread by introducing content to their connections, providing them a direct role in a cascade's ultimate size. This makes those individuals particularly important for applications in which the spread of content needs to be augmented or slowed, such as in marketing or harm mitigation applications. 

However, many point process methods are computationally expensive and unable to scale to very large datasets. Even with recent efforts at optimization, memory scaling remains a critical roadblock for implementing MHPs at scale \citep{nickel2021modeling}. While methods like neural network-based MHPs \citep{mei2017neural} can scale somewhat, their functional limits still lie short of cascade datasets with many entities and event sequences \citep{nickel2021modeling}. For those cases, a scalable individual-level influence proxy is desirable.

This work proposes a novel individual-level metric of cascade participation, which attempts to capture the extent to which each member across a set of cascade is responsible for seeding content to other participants. Rather than attempting to model cascades in their entirety, our approach is predicated on the desire for an approach that identifies disproportionately influential individuals. In our evaluation, we demonstrate that this approach retains valuable individual-level information while decreasing runtime requirements by orders of magnitude relative to MHPs.

\section{Related Work}
Prior work has approached digital information flow from a few perspectives. First, many studies attempt to model the structure of flow, particularly as it occurs across networks. Broadly, this approach encompasses a wide variety of network-based information diffusion models. More narrowly, a line of research in this vein attempts to model viral cascades, instances in which information rapidly spreads along structural paths. Second, there is a line of research that attempts to glean content-level drivers of spread. This research includes attempts to identify specific characteristics of content that induce virality, as well as efforts to model granular content-level popularity over some time period. Finally, a less common approach in this space is examining individual-level contribution to information flow. In this approach, researchers attempt to identify influence within flow, distinguishing people who are disproportionately influential to a piece of content’s virality. 

Our work builds on the third approach, providing a novel way to identify individuals who play a large role in seeding viral cascades. In the following sections, we describe the work done within each approach, highlighting the overlaps and distinctions between each. We then highlight the gaps in these approaches, demonstrating the contribution of our approach to this body of work.

\subsection{Structures of information flow}
Many researchers approach understanding information diffusion as a structural problem. Information passes through connections, making it a natural fit for networked models \citep{bakshy2012role}. Some pathways within those networks are more prolific than others at spreading information \citep{benkler2018network}. Identifying those pathways, and the individuals who participate in them, allows us to better understand the dynamics underpinning how information gets transmitted. 

Much of the work in this space draws from models of epidemic spreading. These models treat information as a disease, which ``infects'' users in a network as they circulate it further \citep{zhou2021survey}. They allow researchers to evaluate key factors in the dynamics of information spread---such as magnitude and rate of spread---and how they respond to variations in network structure, composition, and individual inclination.  

A subset of modeling information diffusion focuses on viral cascades as a way to unpack the structure of spread. Cascades are repeated sequences of actions, around a particular piece of content, that extend that content’s reach to an increasingly larger audience over time \citep{dow2013anatomy}. These can, in social media contexts, take the form of reshares, retweets, or any other feature that enables organic extension of content’s audience. By the nature of how they increase exposure and present opportunities for further spread, cascades can rapidly increase the audience for content by orders of magnitude \citep{dow2013anatomy}.

Work on cascades covers several key aspects. First are efforts to predict when a cascade will form, and how large it will be, using network, temporal, and activity-based characteristics \citep{cheng2014,zhou2021survey}. Other research attempts to elucidate the structure of cascades, and how that structure might vary depending on facets of collective behavior \citep{dow2013anatomy}. This work feeds indirectly into prediction, as some cascade structures appear to have more potential for viral growth than others. Finally, there are attempts to map cascades to their underlying mechanisms, drawing associations between how the cascade progresses and how the users are connected in the organic interaction network \citep{myers2010convexity}. This second-order mapping provides a path toward predicting the likelihood and structure of a cascade, based on how the underlying network structures itself at a given point in time.

Cascade-based approaches have proven successful at predicting information spread in some contexts \citep{cheng2014}. They also elucidate the dynamics of that spread from a structural perspective, helping to make the underlying activity network explicit. However, some areas of cascade-based virality are still relatively underexplored. Chief among these is the idea of decomposing a cascade, acknowledging that some phases or participants are more consequential to spread than others \citep{mathew2019spread}. Some prediction approaches acknowledge the strong relationship between a cascade’s early and late performance \citep{szabo2010predicting}. There is less work exploring the possibility that a cascade’s dynamics can meaningfully hinge on certain individuals or stages throughout the course of its lifecycle. In addition, examinations of cascades are often limited to individual cases, treating cascades in isolation \citep{szabo2010predicting,cheng2014}. It is often useful to understand dynamics across a set of cascades, in order to understand the (lack of) consistency in behavior across content or time. In short, current approaches lack a straightforward way to measure consistent individual contribution across multiple cascades.

\subsection{Measuring individual contribution}
Not every individual who participates in a cascade is equally responsible for its spread. To borrow again from epidemiology, a relatively small number of ``superspreaders'' are responsible for large amounts of infection \citep{pei2014searching}. Some participants are more authoritative than others, and thus better primed to circulate information within a system---this is the logic that motivates PageRank \citep{gleich2015pagerank}. And some participants are better at seeding cascades in such a way that they propagate out to many others \citep{wilson2020cross}. Efforts to identify these consequential individuals fall under a few categories.

First, researchers have attempted to identify users who are well situated to create popular content. One line of research searches for influencers, picking out individuals who hold sway over a community or network \citep{harrigan2021identifying}. Inversely, some researchers seek to identify individuals who are consistently unable to pick popular offerings, thereby providing an indirect mechanism to gauge content or product popularity \citep{anderson2015harbingers}. These attempts don’t necessarily rely on cascades; in many cases, they simply connect people to content and pick out those individuals with the best track record. And while this approach can be effective for straightforward identification, it provides neither a granular view of influence nor an understanding of where in the process of spread these individuals fall. 

Second, there are some efforts to manipulate cascades based on their structures. These approaches identify crucial points at various stages of cascades–what combinations of users, content, and structure affect their formation, for example, or at which points widespread virality is most likely to occur \citep{gruhl2004information}. These approaches are useful for powering interventions. Understanding the crucial components of a cascade can help a system foster them, to encourage increased virality, or stifle them, to limit undesirable content. 

Finally, some approaches attempt to infer an underlying influence structure from the dynamics of cascade participation. Algorithms like NetInf infer an underlying network structure among information spreaders, based on the activity observed in cascades \citep{myers2010convexity}. Point process based approaches, such as Multivariate Hawkes Processes (MHPs), model event sequences by attempting to discern the likelihood of an event at a given point in time \citep{farajtabar2017fake}. In the process of this modeling task, MHPs also produce individual-level measures of who influences whom \citep{nickel2021modeling}. This byproduct, while computationally expensive, provides an interpretable view of which participants across a cascade set exert the most influence on its activity. 

Our work attempts to accomplish a similar goal, without the same magnitude of computational cost. The method we propose here unravels user-level contributions to cascade success, relying only on timestamped activity information. Our goal is to allow for scalable decomposition of cascades into their most consequential participants. 

\section{Method}
The objective of our approach is to capture user-level variation in participation across a set of cascades, such that it becomes straightforward to identify users who play a clear role in the early stages of those cascades. Critically, our approach assumes no knowledge about the underlying structure of cascades. This allows it to operate with minimal information about cascade participation, and to computationally scale to very large datasets. Toward this objective, we rely on three conceptual characteristics of broadly influential users:

First, these users tend to be consistently \textit{early} to cascades. In the language of prior work, they are in the ``ramp up'' period of a cascade's lifecycle, at or near the point where a novel event sequence originates \citep{gruhl2004information}. The most consequential users repeatedly appear in this early stage of a cascade's spread, making them strong candidates for downstream influence. 

Second, these users participate in \textit{viral cascades}. In other words, they are routinely successful at participating in cascades which later get traction from many other participants. In some scenarios it is helpful to distinguish between cascade participants who are widely influential, and those who, e.g., repeatedly spam a piece of content without getting any uptake from others. 

Finally, these users may have \textit{small direct audiences}. In some measures of influence or importance, engagement metrics provide a useful proxy for a participant's popularity \citep{hagar2022concentration}. More knowledge of the structure of information diffusion allows for more nuanced capture of indirect influence, by virtue of association with popular actors \citep{liu2017influence}. However, in the absence of structural information, we cannot assume that a given participant's direct audience accounts for the entirety of their influence. Rather, they may be an early participant in the \textit{local} spread of a cascade, before it reaches a broader group of participants, and only reach a small fraction of individuals directly \citep{wilson2020cross}.

To capture these characteristics, we develop a lightweight individual-level metric over a given set of cascades:

\begin{equation}\label{metric}
\sum_{c=1}^{C}ln(d)\cdot v\cdot p^{\alpha}
\end{equation}

In equation \ref{metric}, \textit{d} represents the number of participants \textit{downstream} of an individual in a cascade (i.e., the number of users who post a piece of content after that user). \textit{v} is an optional binary variable, intended to indicate whether an individual's contribution to the cascade received any attention. This term is helpful for filtering out, e.g., social media posts that were viewed by no one, where that information is available. \textit{p} is the inverse percentile of a user’s position in a cascade (i.e., for the first participant p=1; for the last p=0). The exponent $\alpha$ adds a strong decay, so that participants who are late to very large cascades are not disproportionately rewarded. We have found success setting $\alpha=0.5$, which reduces all values after the halfway point of a cascade to $\approx0$. 

These cascade-level values are summed for all cascades in which an individual participates. An additive metric has several key benefits in practical applications. First, the metric rewards individuals who are repeatedly early to the largest cascades in a given set, providing meaningful differentiation among participants across multiple event sequences. Second, this straightforward calculation makes it possible to decompose an individual's score and determine which cascades contributed most to it. Inversely, in real-world contexts with a continuous stream of cascade data over an extended period of time, it is possible to continuously track how individuals' influence grows without revisiting earlier data. Finally, we can produce an inferred influence network among cascade participants, using this metric as an edge weight. This is analogous to the network inference performed by many other influence measurement approaches \citep{myers2010convexity}. However, unlike many of those methodologies, ours does not require information about the preexisting network structure. 

\subsection{Comparison to existing methods}
Because it is concerned with quantifying cascade influence at the individual level, our approach can be most directly compared to other cascade modeling techniques. In this section, we specifically make comparisons to three widely-used cascade modeling approaches: \textit{feature-based prediction}, \textit{Multivariate Hawkes Processes}, and \textit{network inference algorithms}. Table \ref{comparison} summarizes key points of comparison across methods. 

\begin{adjustbox}{angle=90,totalheight=7in,caption={Comparison of approaches to analyzing cascade data},float=table}
    \begin{tabular}{@{}llllll@{}}
    \toprule
    \textbf{Method}          & \textbf{Parameters to learn}          & \textbf{Feature engineering required} & \textbf{Input data}             & \textbf{Scaling constraints} & \textbf{Output(s)}                                                         \\ \midrule
    Early adopter metric     & alpha                                 &                                       & Time-ordered event sequences    & Cascade generation           & Individual-level early adoption score                                      \\
    MHPs                     & kernel estimation, baseline intensity &                                       & Time-ordered event sequences    & Event and participant set    & Model of underlying event generation, inter-participant influence measures \\
    NetInf                   & cascade spreading likelihoods         &                                       & Sequences and network structure & Network size                 & Inferred influence network structure                                       \\
    Feature-based prediction & model dependent                       & X                                     & Engineered feature set          & Feature complexity           & Cascade growth prediction                                                  \\ \bottomrule
    \end{tabular}
    \label{comparison}
\end{adjustbox}

\subsubsection{Feature-based prediction}
Feature-based cascade prediction differs from our approach in its goal, which is often to determine the size of a cascade at some time $t+n$ from the information available at time $t$ \citep{cheng2014}. However, such approaches sometimes leverage individual-level features in their prediction tasks, including by attempting to identify disproportionately influential participants \citep{hagar2022concentration}. Depending on the predictive model used for this task, it is even possible to draw a connection between (subsets of) cascade participants and the model's ability to predict cascade size, producing a similar individual-level measure of importance to ours. 

The key differences between the approaches stem from a difference in desired outcomes. Because this approach to cascade modeling involves a more complex prediction task, it requires sophisticated models with their own parameters to learn. This approach requires explicit training for the model, necessitating a large dataset from which to learn associations. It also relies on costly feature engineering, and the assumption that the features selected by the modeler will be predictive of the desired outcome \citep{zhao2015seismic}.

In contrast, our approach requires no model training. Individual-level importance measures are the sole outcome, rather than a byproduct of a larger modeling task. Our approach also does not require feature engineering, beyond encoding the time-ordered event sequences themselves. 

\subsubsection{Multivariate Hawkes Processes}
Across the board, our approach shares the most similarity with Multivariate Hawkes Processes (MHPs). There are a variety of approaches to training MHPs, and there are a number of extensions and related methods that take advantage of the underlying MHP approach to some extent \citep{nickel2021modeling,mei2017neural}. This comparison considers the baseline implementation of MHPs. Both approaches produce individual-level measures of influence. Both infer an influence structure from a set of time-ordered event sequences, and both require only that sequence information to operate. However, our approach prioritizes different trade offs than those required by MHPs. 

MHPs provide a fuller picture of the structure and dynamics of influence underpinning an event set. They allow researchers to model the generative process behind an observed sample \citep{embrechts2011multivariate}. Our approach makes no attempt to glean information about the cascades in a sample themselves. Our approach's scoring method relies largely on position, rather than inferred inter-participant interaction. MHPs do attempt to infer this interaction structure, in addition to a baseline event probability, providing a more detailed view of how influence might flow among participants \citep{rambaldi2017role}.

This detailed model comes at the expense of severe computational limitations, though. MHPs often cannot scale past networks with hundreds or thousands of nodes. Even after careful optimization, such models are hampered by sample size \citep{nickel2021modeling}. Our approach, on the other hand, has no such computational constraint. The main barrier to efficient calculation for our metric stems from the ability to generate underlying cascade data, rather than from the metric itself. This makes it straightforward to apply to massive pre-existing datasets, without meaningful added computational overhead. 

\subsubsection{Network inference}
A related problem to the tasks explored here is network inference: In the absence of explicit (often impossible to collect) network structure data, can we infer a network from a dataset of activity? Network inference is not limited to cascade data; it can, for example, draw from more general models of information diffusion or disease contagion \citep{gomez2012inferring}. For this comparison, we focus on NetInf \citep{gomez2012inferring}. Variations and extensions of this algorithm have been proposed but rely on similar underlying principles \citep{rodriguez2011uncovering,myers2010convexity}. 

NetInf attempts to infer the network structure of influence from a sample of activity. Similar to MHPs, it relies solely on timestamped event sequences to generate its output. Unlike the other methods covered here, NetInf is focused on modeling network structure. It takes a probabilistic approach, aimed at determining the most likely directed network corresponding to observed activity \citep{gomez2012inferring}. 

As a result, NetInf does not directly produce any individual-level influence measure within the larger structure. The directed network it infers can implicitly gauge influence---\citet{gomez2012inferring} leverage the inferred hub-and-spoke structure of news dissemination to identify key outlets. However, this represents an application of standard network analysis techniques to the implied structure after the fact, rather than a facet of the modeling approach.

\subsubsection{Comparison summary}
Compared to our approach, the methods summarized here represent more complex attempts to model some aspect of diffusion or activity. All address related problems, but only MHPs overlap in explicitly producing individual-level measures of influence within a cascade sample. And relative to MHPs, our method does not suffer from meaningful computational constraints at scale, making it well-suited for large datasets. This situates our method as a scalable attempt at individual-level influence inference, expressed through an interpretable metric and independent of any broader modeling task. 

\subsection{Implementation considerations}
There are two main practical considerations for implementations of this method. First is the need for clear understanding of the distribution of participant count in the cascade sample. Because downstream participant count factors heavily into the metric, large cascades count for more than small ones. An individual who is early to several very large cascades, then, will score higher than one who is early to many small ones. This characteristic may not be desirable depending on the type of individual being identified in a particular use case, and can be somewhat alleviated by placing constraints on cascade size for inclusion in a sample. 

Second, implementations should consider the balance between consistency, position, and size this metric captures. Scores may not be comparable across cascade sets, and interpreting the drivers of an individual's score is infeasible without an understanding of the underlying data. 

\subsection{Scope and limitations}
As mentioned above, our metric is a focused attempt to glean participant-level information from cascade data. Specifically, it is designed to capture \textit{early adopters}---individuals who are consistently among the first to participate across a sample of cascades. This scope introduces two important limitations. First, we do not consider other conceptual models of individual influence or importance. A participant could be relatively late to a cascade, but influential by virtue of their large audience. Or, they could exercise influence measurable not by its magnitude, but by its efficacy in achieving some desired outcome outside of the observed data. Our approach does not consider these cases, as it hews to a framework of influence informed by adoption order. 

Second, as with many of the methods explored above, ours relies solely on timeseries data. It does not account for the actual cascade structure underpinning the observed events. That structural information is often not available, but where it is, researchers have observed tree-like patterns of diffusion with critical paths \citep{cheng2014}. By only considering temporal information, our method runs the risk of identifying dead-end ``leaves'' on that tree as highly influential, rather than distinguishing participants on the critical paths. 

\section{Experiments}
In the following, we demonstrate our method on several real-world datasets to demonstrate three key properties: 1) interpretable individual-level measures of cascade influence, 2) scalable calculation, both in runtime and memory usage, and 3) extensions that enable real-time score updating over repeated observations. 

\subsection{Influence scores in policy diffusion}
In this analysis, we demonstrate the straightforward interpretability our measure provides to distinguish disproportionately influential cascade participants. We rely on the State Policy Innovation and Diffusion Database (SPID) \citep{boehmke2020spid}. This database, collected to study state-level policy diffusion in the U.S., contains 724 policies implemented between 1691 and 2017. The database records which states adopted each of these policies, and in which year. 

We treat each policy in this dataset as a cascade, and each state as a cascade participant, ordered by the year that they adopted each policy. This gives us a sample of 724 cascades, ranging in size from 1 to 50 participants (median=23). From this sample, we calculate our early adopter metric for each state, and rank states by their scores. 

Table \ref{tab:spid} shows the top and bottom five states, ordered by their early adopter scores. The states we identify as the most influential generally align with prior work. \citet{nickel2021modeling} also highlight New York and California as two of the most globally influential states. And in their paper introducing this database, \citet{boehmke2020spid} identify all five of our highest ranked states as among the most consequential to policy diffusion.

\begin{table}[!ht]
\begin{tabular}{@{}lll@{}}
\toprule
\textbf{Rank} & \textbf{State} & \textbf{Score} \\ \midrule
1             & California     & 532.1          \\
2             & Massachusetts  & 428.7          \\
3             & New York       & 420.6          \\
4             & Connecticut    & 416.9          \\
5             & Minnesota      & 407.9          \\
...           & ...            & ...            \\
46            & Georgia        & 209.4          \\
47            & Alabama        & 206.0          \\
48            & Wyoming        & 170.0          \\
49            & Mississippi    & 147.3          \\
50            & Alaska         & 145.3          \\ \bottomrule
\end{tabular}
\caption{Scores generated by our calculation, for the top and bottom five states by inferred influence. These rankings align roughly with those generated in prior work.}
\label{tab:spid}
\end{table}

To understand how we arrived at these scores, we can also examine each state's relative performance in each of our metric's terms. Doing so reveals interesting variation in behavior---Hawaii, for example, is a relatively early but infrequent participant; Washington adopts many policies that get relatively little adoption from other states, and Mississippi is a consistent late adopter to popular policies. These details might help researchers distinguish among different \textit{types} of influence, or to further understand why certain cascade participants scored higher than others. 

\subsection{Runtime on Memetracker data}
As mentioned above, one of the primary objectives of our approach is to provide a lightweight metric that scales to very large datasets. MHPs have historically been limited by computational constraints, and even optimized approaches can take hours to process samples with hundreds of thousands of events \citep{nickel2021modeling}. In contrast, our approach can run on datasets with millions of events, in a fraction of the time, on consumer-grade hardware.

To illustrate, we draw on a subset of the Memetracker dataset \citep{gomez2012inferring}. We compare the runtime of our calculation to the values reported by \citet{nickel2021modeling} in evaluating several MHP-based approaches on subsets of the same dataset. In their experiments, the authors test standard MHPs, Neural Hawkes Processes \citep{mei2017neural}, and their own sparse MHP optimization on four subsets of the full Memetracker dataset. The subsets range in size from 64,000 to 400,000 events. On enterprise-grade hardware, \citet{nickel2021modeling} report wall clock times for a single training epoch ranging from a few seconds (for lazy MHPs) to 5+ hours (for standard MHPs). 

We run tests on the same subsets, reported in table \ref{tab:runtimes}. The full runtime for our algorithm ranges from 1.14 to 6.05 seconds on these subsets, in line with a single epoch of lazy MHP training.

\begin{table}[!ht]
\begin{tabular}{@{}ll@{}}
\toprule
\textbf{Cascade set} & \textbf{Mean runtime (seconds)} \\ \midrule
Amy Winehouse        & 3.60                            \\
Arab Spring          & 6.05                            \\
bail out             & 1.14                            \\
Miami Heat           & 1.93                            \\ \bottomrule
\end{tabular}
\label{tab:runtimes}
\caption{Mean runtime of our approach on Memetracker subsets across seven iterations}
\end{table}

These results demonstrate the crux of our tradeoff---while our approach does not generate detailed modeling of underlying processes, it can rapidly provide individual-level influence scoring at scale. 

\subsection{Extension: Online influence scores}
In all the examples presented from the existing methods explored here, analyses focused on post-hoc modeling of full cascades. Historical data is a necessity for inferring influence structure, as truncated samples obscure potentially critical influence relationships. At the same time, the assumption that samples of full cascades are available often does not hold in real-world data. Or, more precisely, real-world cascades often lack a clear endpoint, making it uncertain when analysis should proceed. Consider the case of social media virality, where a URL might spread among users for a few hours, vanish, and then reappear months later as it resurges in popularity. Such cases necessitate an approach that is flexible enough to recalculate influence measures over multiple time periods, and to do so while avoiding computational redundancy and retaining the history of prior calculations. 

To demonstrate how our method meets these criteria, we return to the Memetracker sample used above. This dataset provides a retrospective view of information spreading. For the purposes of this analysis, we break the data into 20 evenly-spaced samples over time, containing between 12,000 and 120,000 observations each. Because our metric is \textit{additive}, it is trivial to 1) maintain a running outlet-level influence score from available historical data, and 2) update that score as a new month's data become available.

\begin{figure}
    \centering
    \includegraphics[width=\textwidth]{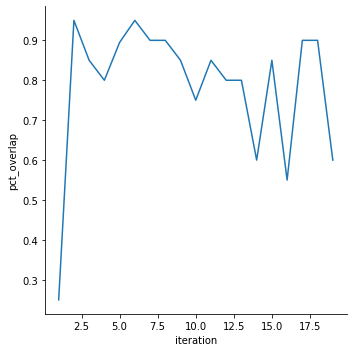}
    \caption{Percent of the top twenty participants by influence score that remain consistent from interval to interval, based on a rolling 3-interval window.}
    \label{fig:consistency}
\end{figure}

To demonstrate, we calculate participants' influence scores on a rolling 3-interval basis across the sample. We then rank participants by their score within each window, then calculate how consistent the top 20 participants remain over each interval. Figure \ref{fig:consistency} visualizes the relative consistency in these rankings over the full time window. 

This view reveals dynamics that are less clear in analyses that attempt to model entire cascades after their completion---influence changes hands over the course of a cascade's lifetime, and some influential actors only appear as new cascades develop later in the time window. This approach may aid practitioners in monitoring the development of novel cascades, and in measuring how the actors seeding those cascades differ from others in the sample. 

\section{Conclusion}
In this work, we have developed a novel approach for deriving participant-level measures of influence from multi-cascade samples. By focusing on decomposing cascade participation by timeliness, size, and indirect downstream impact, we are able to precisely identify individuals who are disproportionately responsible for widespread virality. Our approach can accomplish this at larger scale and with far fewer computational resources than other state-of-the-art influence modeling methods, such as MHPs. 

This approach opens up important avenues for examining the \textit{dynamics} of cascade spread. It presents an opportunity to test theorized differentiation in the role of individual cascade participants \citep{gruhl2004information}, and to examine how influence shifts over long time windows. For practical applications, it provides a lightweight, additive metric for ongoing influence monitoring, allowing straightforward delineation of a population of users into its most and least consequential drivers of viral cascades.

\bibliographystyle{apalike}
\bibliography{references}

\end{document}